\documentstyle[pre,aps]{revtex}
\def\H{{\mathcal{H}}}
\def\I{{\mathcal{I}}}
\def\Y{{\mathcal{Y}}}
\begin{document}
\draft
\title{Information Theory based on Non-additive Information Content}
\author{Takuya Yamano}
\address{Department of Applied Physics, Faculty of Science, Tokyo Institute of 
Technology, Oh-okayama, Meguro-ku, Tokyo,152-8551, Japan}
\maketitle
\begin{abstract}
We generalize the Shannon's information theory in a nonadditive way by focusing on the source 
coding theorem. The nonadditive information content we adopted is consistent with the concept 
of the form invariance structure of the nonextensive entropy. Some general properties of 
the nonadditive information entropy are studied, in addition, the relation between 
the nonadditivity $q$ and the codeword length is pointed out.  
\end{abstract}

\section{Introduction}
 
The intuitive notion of what a quantitative expression for information should be 
has been addressed in the development of transmission of information which led to 
the information theory(IT). The IT today is considered to be the most fundamental 
field which connects other various fields such as physics 
(thermodynamics), electrical engineering (communication theory), mathematics 
(probability theory and statistics), computer science (Kolmogorov complexity) 
and so on\cite{Cover}. Accordingly, the selection of the information measure 
becomes a influential matter.
The introduction of logarithmic form of information measure dates back to 
Hartley\cite{Hartley}. He defined the practical measure of information as the 
logarithm of the number of possible symbol sequences. After that, Shannon established the 
logarithmic based IT from the reasons: (a) practical usefulness, (b) 
closeness to our intuitive feeling, (c) easiness of mathematical 
manipulation\cite{Shan,ShanWea}.\\

On the other hand, however, non-logarithmic form of (or nonextensive) entropy is currently  
considered as a useful measure in describing thermostatistical properties of a certain class of 
physical systems which entail long-range interactions, long-time memories and (multi)fractal structures. The form of the nonextensive entropy proposed by Tsallis\cite{88Tsallis} has been 
intensively applied to many such systems\cite{biblio}. The reason why the formalism  
violating the extensivity of the statistical mechanical entropy seems to be essential for 
convincing description of these systems is not sufficiently revealed in the present status.
Nevertheless the successful application to some physical systems seems to lead us to investigate
into the possibility of the nonadditive IT since Shannon's information entropy has the same 
form as the logarithmic statistical mechanical entropy.\\

It is desirable to employ the nonadditive information content which the 
associated IT contains Shannon's IT in a special case. The concept of the form 
invariance to the structure of nonextensive entropies was considered 
to provide a guiding principle for a clear basis for generalizations of 
logarithmic entropy\cite{RajAbe}. This structure seems to give a hint at the selection of 
the nonadditive information content. The form invariant structure require 
normalization of the original Tsallis entropy by $\sum_ip_i^q$, where 
$p_i$ is a probability of 
event $i$ and $q$ is a real number residing in the interval $(0,1)$ from the 
preservation of concavity of the entropy. In addition, Kullback-Leibler (KL) 
relative entropy which measures distance between two probability distribution 
is also modified\cite{RajAbe}.\\

This paper explores consequences of adopting nonadditive information content 
in the sense that the associated average information i.e entropy takes a form 
of the modified Tsallis entropy. The use of modified form of Tsallis entropy 
is in conformity with the appropriate definitions of expectation value (the 
normalized $q$-expectation value\cite{98Tsallis}) of the nonadditive information content. 
Since the information theoretical entropy is defined as the \textit{average} of information 
content, it is desirable to unify the meaning of the \textit{average} as the 
normalized $q$-expectation value throughout the nonadditive IT.
Moreover we shall later see how the Shannon's additive IT is extended to the 
nonadditive one by addressing the source coding theorem which is the one of the 
most fundamental theorem in IT.\\

The organization of this paper is as follows. In Sec.II, we present the 
mathematical preliminaries of the nonadditive entropy and the generalized KL 
entropy. Sec.III addresses an optimal code word within the framework of 
nonadditive context. We shall attempt to give a possible meaning of 
nonadditive index $q$ in terms of codeword length there. Sec.IV deals with the 
extension of the Fano's inequality which gives upper bound on the conditional 
entropy with an error probability in a channel.
In the last section, we devote to concluding remarks.

\section{Nonadditive entropy and the generalized KL entropy}
\subsection{Properties of nonadditive entropy of information}
For a discrete set of states with probability mass function $p(x)$, 
where $x$ belongs to alphabet $\H$,  
we consider the following nonadditive information content $I_q(p)$,
\begin{equation}
I_q(p)\equiv -\ln_q p(x)\label{eqn:NIC},
\end{equation}
where $\ln_q x$ is a $q-$logarithm function defined as 
$\ln_q x = (x^{1-q}-1)/(1-q)$. In the limit $q \rightarrow 1$, $\ln_q x$ 
recovers the standard logarithm $\ln x$. In Shannon's additive IT, an 
information content is expressed as $-\ln p(x)$ in \textit{nat} unit, which is monotonically 
decreasing function with respect to $p(x)$. This behavior matches our
intuition in that we get more information in the case the least-probable 
event occurs and less information in the case the event with high probability 
occurs. It is worth noting that this property is qualitatively valid for nonadditive 
information content for all $q$ except the fact that there exists upper 
limit $1/(1-q)$ at $p(x)=0$. Therefore the Shannon's reason (b)
we referred in Sec.I is considered to be no crucial element for 
determining the logarithmic form. Moreover, it is easy to see that the Renyi information of 
order $q$\cite{Renyi}, $H_q^R=\displaystyle{\ln\sum_{x\in \H}}p^q(x)/(1-q)$, which is additive information measure can be written with this nonadditive 
information content as
\begin{equation}
H_q^R=\frac{\displaystyle{\ln\sum_{x\in \H}}\left[ 1-(1-q)I_q(p(x))\right]^{\frac{q}{1-q}}}
{1-q}.
\end{equation}

 The entropy $H_q(X)$ of a discrete random variable $X$ is defined as an average of 
the information content, where the \textit{average} means the normalized $q$-expectation value
\cite{98Tsallis},
\begin{equation}
H_q(X) = \frac{\displaystyle{-\sum_{x\in \H}}p^q(x)\ln_qp(x)}
{\displaystyle{\sum_{x\in \H}}p^q(x)} = \frac{1-\displaystyle{\sum_{x\in \H}}
p^q(x)}{(q-1)\displaystyle{\sum_{x\in \H}}p^q(x)}\label{eqn:Hx},
\end{equation}
where we have used the normalization of probability $\sum_{x\in \H}p(x)=1$. 
In a similar way, we define the nonadditive conditional information content and the joint one 
as follows
\begin{equation}
I_q(y\mid x) = \frac{p^{1-q}(y\mid x)-1}{q-1},
\end{equation}
\begin{equation}
I_q(x,y) = \frac{p^{1-q}(x,y)-1}{q-1},
\end{equation} 
where $y$ belongs to a different alphabet $\Y$. Corresponding entropy conditioned by $x$ and the joint entropy of $X$ and $Y$ becomes
\begin{equation}
H_q(Y\mid x) = \frac{\displaystyle{\sum_{y\in \Y}}p^q(y\mid x)I_q(y\mid x)}
{\displaystyle{\sum_{y\in \Y}}p^q(y\mid x)}
= \frac{1-\displaystyle{\sum_{y\in \Y}}p^q(y\mid x)}
{(q-1)\displaystyle{\sum_{y\in \Y}}p^q(y\mid x)}\label{eqn:HYx}
\end{equation}
and 
\begin{equation}
H_q(X,Y) = \frac{\displaystyle{\sum_{x\in \H,y\in \Y}}p^q(x,y)I_q(x,y)}
{\displaystyle{\sum_{x\in \H,y\in \Y}}p^q(x,y)}
= \frac{1-\displaystyle{\sum_{x\in \H,y\in \Y}}p^q(x,y)}
{(q-1)\displaystyle{\sum_{x\in \H,y\in \Y}}p^q(x,y)}\label{eqn:Hxy},
\end{equation}
respectively.  Then we have the following theorem.\\
\textbf{Theorem 1.}\\
The joint entropy satisfies 
\begin{equation}
H_q(X,Y)=H_q(X)+H_q(Y\mid X)+(q-1)H_q(X)H_q(Y\mid X)\label{eqn:pseud}.
\end{equation}
\textbf{Proof.}\\
From Eq(\ref{eqn:Hxy}) we can rewrite $H_q(X,Y)$ with the relation 
$p(x,y)=p(x)p(y\mid x)$ as 
\begin{equation}
H_q(X,Y)=\frac{1}{q-1}\left[ \frac{1}{\displaystyle{\sum_{x\in \H}p^q(x)
\sum_{y\in \Y}p^q(y\mid x)}}-1\right].
\end{equation}
Since Eq(\ref{eqn:HYx}) gives $\displaystyle{\sum_{y\in \Y}p^q(y\mid x)}=
\left[ 1+(q-1)H_q(Y\mid x)\right]^{-1}$, we get 
\begin{equation}
H_q(X,Y)=\frac{1}{q-1}\left[ \frac{1}{\displaystyle{\sum_{x\in \H}}\frac{p^q(x)}
{1+(q-1)H_q(Y\mid x)}}-1\right].\label{eqn:HXYj}
\end{equation}
Here, we introduce the following definition\cite{defofcond}
\begin{equation}
\langle \frac{1}{1+(q-1)H_q(Y\mid x)}\rangle^{(X)}_q \equiv \frac{1}{1+(q-1)H_q(Y\mid X)}\label
{eqn:defcond}
\end{equation}
where the bracket denotes the normalized $q$-expectation value with respect to $p(x)$.
Then we have from Eq(\ref{eqn:HXYj}), 
\begin{equation}
H_q(X,Y)=\frac{1}{q-1}\left[ \frac{1+(q-1)H_q(Y\mid X)}{\displaystyle{\sum_{x\in \H}}p^q(x)} 
-1\right].
\end{equation}
Putting $\displaystyle{\sum_{x\in \H}}p^q(x) = \left[ 1+(q-1)H_q(X)\right]^{-1}$ 
into this yields the theorem.$\Box$\\
This theorem has a remarkable similarity to the relation with which the Jackson basic number 
in $q$-deformation theory satisfies, which was pointed out in Ref.\cite{RajAbe}. 
That is, $[X]_q\equiv (q^X-1)/(q-1)$ is the Jackson basic number of a 
quantity $X$. Then, for the sum of two quantities $X$ and $Y$, the assosiated 
basic number $[X+Y]_q$ is shown to become $[X]_q+[Y]_q+(q-1)[X]_q[Y]_q$.
Obviously this theorem recovers ordinary relation $H(X,Y)=H(X)+H(Y\mid X)$ 
in the limit $q\rightarrow 1$.
In this modified Tsallis formalism, $q$ appears as $q-1$ instead of as 
$1-q$\cite{AbeRaj1,Abe}. When $X$ and $Y$ are independent events each other, 
Eq(\ref{eqn:pseud}) gives the pseudoadditivity relation\cite{98Tsallis}. However, 
it is converse to the case of the original Tsallis one in that $q > 1$ yields 
superadditivity and $q < 1$ subadditivity.
It is worth mentioning that the concept of nonextensive conditional entropy in the framework 
of the original Tsallis entropy has firstly introduced for discussing quantum entanglement 
in Ref.\cite{AbeRaj1}. From this theorem, we immediately have the following corollary 
concerning the equivocation.\\
\textbf{Corollary.}
\begin{equation}
H_q(Y\mid X)=\frac{H_q(Y,Z\mid X)-H_q(Z\mid Y,X)+(q-1)
\{ H_q(X)H_q(Y,Z\mid X)-H_q(X,Y)H_q(Z\mid Y,X)\}}{1+(q-1)H_q(X)}\label{eqn:Cor}
\end{equation} 
\textbf{Proof.}\\
In Eq(\ref{eqn:pseud}), when we see $Y$ as $Y,Z$ we have
\begin{equation}
H_q(X,Y,Z)=H_q(X)+H_q(Y,Z\mid X)+(q-1)H_q(X)H_q(Y,Z\mid X)
\label{eqn:YYZ},
\end{equation}
on the other hand, when we regard $X$ as $Y,X$ and $Y$ as $Z$, we get 
\begin{equation}
H_q(X,Y,Z)=H_q(X,Y)+H_q(Y,Z\mid X)+(q-1)H_q(X,Y)H_q(Z\mid Y,X)
\label{eqn:XYZ}.
\end{equation}
Subtracting the both sides of the above two equations and arranging with respect 
to $H_q(Y\mid X)$ with Eq(\ref{eqn:pseud}), we obtain the corollary.
$\Box$\\
Remarks: In the additive limit($q\rightarrow 1$), we recover 
the relation  $H(Y\mid X) = H(Y,Z\mid X) - H(Z\mid Y,X)$.\\
Moreover, with the help of Eq(\ref{eqn:Cor}), we have the following theorem.\\
\textbf{Theorem 2.} Hierarchical structure of entropy $H_q$\\
The joint entropy of $n$ random variables $X_1,X_2,\ldots ,X_n$ satisfies
\begin{equation}
H_q(X_1,X_2,\cdots ,X_n)=\sum_{i=1}^n \left[ 1+(q-1)H_q(X_{i-1},\cdots ,
X_1)\right]H_q(X_i\mid X_{i-1},\cdots ,X_1)\label{eqn:chain}.
\end{equation}
\textbf{Proof.}\\
From Eq(\ref{eqn:pseud}), $H_q(X_1,X_2) = H_q(X_1) + \left[ 1+(q-1)H_q(X_1)
\right]H_q(X_2\mid X_1)$ holds. Next, from Eq(\ref{eqn:YYZ}), we have 
\begin{equation}
H_q(X_1,X_2,X_3)=H_q(X_1)+H_q(X_2,X_3\mid X_1)+(q-1)H_q(X_1)H_q(X_2,X_3\mid X_1)
\label{eqn:X123}.
\end{equation}
Since $H_q(X_2,X_3\mid X_1)$ is written using Eq(\ref{eqn:Cor}) as
\begin{equation}
H_q(X_2,X_3\mid X_1)=H_q(X_2\mid X_1)+\frac{1+(q-1)H_q(X_1,X_2)}
{1+(q-1)H_q(X_1)}H_q(X_3\mid X_2,X_1),
\end{equation} 
Eq(\ref{eqn:X123}) can be rewritten as 
\begin{equation}
H_q(X_1,X_2,X_3)=H_q(X_1)+\left[ 1+(q-1)H_q(X_1)\right]H_q(X_2\mid X_1)
+\left[ 1+(q-1)H_q(X_1,X_2)\right]H_q(X_3\mid X_2,X_1).
\end{equation}
Similarly, repeating application of the corollary gives the theorem.
$\Box$\\
Remark: In the additive limit, we get $H(X_1,X_2,\cdots ,X_n)=
\sum_{i=1}^n H(X_i\mid X_{i-1},\cdots ,X_1)$ which states 
that the entropy of $n$ variables is constituted by the sum of the conditional 
entropies (\textit{Chain rule}).\\ 
From this relation Eq(\ref{eqn:chain}), we need all joint entropy below the level of 
$n$ random variables to acquire the joint entropy $H_q(X_1,\ldots ,X_n)$, which the 
situation is similar to the BBGKY hierarchy in the $N$-body distribution function.  
Let us next define the mutual information $\I_q(Y;X)$, which quantifies 
the amount of information that can be gained from one event $X$ about another event $Y$,
\begin{equation}
\I_q(Y;X) \equiv H_q(Y) - H_q(Y\mid X) = \frac{H_q(X)+H_q(Y)-H_q(X,Y)+(q-1)H_q(X)H_q(Y)}
{1+(q-1)H_q(X)}.
\end{equation}
Therefore $\I_q(Y;X)$ expresses the reduction in the uncertainty of $Y$ due to 
the acquisition of knowledge of $X$. Here we postulate that the mutual information in 
nonadditive case is non-negative. The non-negativity may be considered to be a requirement
rather than the one to be proved in order to be in consistent with the usual additive 
mutual information.
$\I_q(Y;X)$ also converges to the usual mutual information 
$\I(Y;X)=H(Y)-H(Y\mid X)=H(X)+H(Y)-H(X,Y)$ in the additive case($q\rightarrow 1$). We note 
that the mutual information of a random variable with itself is the entropy itself 
$\I_q(X;X)=H_q(X)$. When $X$ and $Y$ are independent variables, we have 
$\I_q(Y;X)=0$\cite{check}. Then, we have the following theorem.\\
\textbf{Theorem 3.} Independence bound on entropy $H_q$\\
\begin{equation}
H_q(X_1,X_2,\ldots ,X_n) \le \sum_{i=1}^n\left[1+(q-1)H_q(X_{i-1},\ldots ,X_1)\right]
H_q(X_i)
\end{equation}
with equality if and only if each $X_i$ is independent.\\
\textbf{Proof.}\\
From the assumption of $\I_q(X;Y)\ge 0$ introduced above, we have
\begin{equation}
\sum_{i=1}^nH_q(X_i\mid X_{i-1},\ldots ,X_1)\le \sum_{i=1}^nH_q(X_i)
\end{equation}
with equality if and only if each $X_i$ is independent of $X_{i-1},\ldots ,X_1$. Then 
the theorem holds from the previous theorem Eq(\ref{eqn:chain}).
$\Box$\\

\subsection{The generalized KL entropy}
The KL entropy or the relative entropy is a measure of the distance between two 
probability distributions $p_i(x)$ and $p_i'(x)$. Here, we define it as the normalized $q$-expectation value of the change of the nonadditive information content $\Delta I_q\equiv 
I_q(p'(x))-I_q(p(x))$\cite{KL},
\begin{equation}
D_q(p(x)\parallel p'(x))\equiv \frac{\displaystyle{\sum_{x\in \H}}p^q(x)\Delta I_q}
{\displaystyle{\sum_{x\in \H}}p^q(x)}= \frac{\displaystyle{\sum_{x\in \H}}p^q(x)
\left( \ln_qp(x)
-\ln_qp^{\prime}(x)\right)}{\displaystyle{\sum_{x\in \H}}p^q(x)}\label{eqn:Dq}.
\end{equation}
We note that the above generalized KL entropy satisfies the form invariant structure 
which has introduced in Ref.\cite{RajAbe,qlog}.
 We review the positivity of the generalized KL entropy in case of $q>0$ which can be  
considered to be a necessary property to develop the IT.\\
\textbf{Theorem 4.} Information inequality\\
For $q>0$, we have 
\begin{equation}
D_q(p(x)\parallel p'(x))\ge 0
\end{equation}
with equality if and only if $p(x)=p'(x)$ for all $x\in \H$.\\
\textbf{Proof.}\\
The outline of the proof is the same as the one in Ref.\cite{98GKL,Borland} except for 
the factor $\sum_{x\in \H}p^q(x)$. From the definition Eq(\ref{eqn:Dq}),
\begin{eqnarray}
D_q(p(x)\parallel p'(x)) & = & \frac{1}{1-q}\sum_{x\in \H}p(x)\left\{ 1-\left( \frac{p'(x)}{p(x)}
\right)^{1-q}\right\}\Big/\sum_{x\in \H}p^q(x)\\
& \ge & \frac{1}{1-q}\left\{ 1-\left( \sum_{x\in \H}p(x)\frac{p'(x)}{p(x)}\right)^{1-q} \right\}\Big/\sum_{x\in \H}p^q(x)=0
\end{eqnarray}
where Jensen's inequality for the convex function has been used : 
$\sum_xp(x)f(x)\ge f\left(\sum_xp(x)x\right)$ with $f(x)=-\ln_q(x)$, $f''(x)>0$. 
We have equality in the second line if and only if $p'(x)/p(x)=1$ for all $x$, 
accordingly $D_q(p(x)\parallel p'(x))=0$.
$\Box$\\
\section{Source Coding Theorem}
Having presented some properties of the nonadditive entropy and the generalized 
KL entropy as a preliminary, 
we are now in a status to address our main results that the Shannon's source coding theorem can be extended to the nonadditive case. Let us consider encoding the sequence of source 
letters generated from an information source to the sequence of binary codewords as an 
example. If a code is allocated for four source letters $X_1,X_2,X_3,X_4$ as $0,10,110,111$ 
respectively, 
the source sequence $X_2X_4X_3X_2$ is coded into $1011111010$. On the other hand, 
if another code assigns them as $0,1,00,11$, then the codeword becomes $111001$. The 
difference between the two codes is striking in the coding. In the former codeword, 
the first two letters $10$ corresponds to $X_2$ and not the beginning of any other codeword
, then we can observe $X_2$. Next there are no codeword corresponding to $1$ and $11$ but 
$111$ is, and is decoded into $X_4$. Then the next $110$ is decoded into $X_3$, leaving 
$10$ which is decoded into $X_2$. Therefore we can uniquely decode the codeword. In the 
latter case, however, we have possibilities to interpret the first three letters $111$ 
as $X_2X_2X_2$, $X_2X_4$ or as $X_4X_2$. Namely, this code cannot be uniquely decoded 
into the source letter that gave rise to it. Accordingly, we need to deal with so-called 
the \textit{prefix code} 
or the \textit{instantaneous code} such as the former case. The prefix code is a code which 
no codeword is a prefix of any 
other codeword(prefix condition code)\cite{Cover,Gal}. We recall that any code which 
satisfies the prefix condition over the alphabet size $D$($D=2$ is a binary case) must 
satisfy the Kraft inequality\cite{Cover,Gal},
\begin{equation}
\sum_i^MD^{-l_i}\le 1\label{eqn:Kraft}
\end{equation}
where $l_i$ is a code length of $i$th codeword($i=1,\ldots M$).
Moreover if a code is uniquely decodable, the Kraft inequality holds for it\cite{Cover,Gal}.
We usually hope to encode the sequence 
of source letters to the sequence of codewords 
as short as possible, that is our problem is finding a prefix condition code 
with the minimum average length of a set of codewords$\{l_i\}$. The optimal code 
is given by minimizing the following functional constrained by the Kraft inequality,
\begin{equation}
J=\frac{\sum_ip_i^ql_i}{\sum_ip_i^q}+\lambda \left( \sum_iD^{-l_i}\right)
\end{equation}
where $p_i$ is the probability of realization of the word length $l_i$ and $\lambda$ 
is a Lagrange multiplier. We have assumed equality in the Kraft 
inequality and have neglected the integer 
constraint on $l_i$. Differentiating with respect to $l_i$ and 
setting the derivative to $0$ yields 
\begin{equation}
D^{-l_i}=\frac{p_i^q}{\left( \sum_ip_i^q\right)\lambda \log D}.
\end{equation}
Here, it is worth noting that from the Kraft inequality the Lagrange multiplier 
$\lambda$ is related as $\lambda \ge (\log D)^{-1}$. Furthermore when the 
equality holds, the fraction $D^{-l_i^*}$ which is given by the optimal codeword 
length $l_i^*$ is expressed as
\begin{equation}
D^{-l_i^*}=\frac{p_i^q}{\sum_ip_i^q}\label{eqn:Dli}.
\end{equation}
Therefore $l_i^*$ can be written as $\log_D(\sum_ip_i^q)-q\log_Dp_i$ and in the additive 
limit, we obtain $l_i^*=-\log_Dp_i$. However, we can not always determine the optimal 
codeword length like this since the $l_i$'s  must be integers. We have the following theorem.\\
\textbf{Theorem 5.}\\
The average codeword length $\langle L\rangle_q$ of any prefix code for a random variable $X$
satisfies
\begin{equation}
\langle L \rangle_q \ge H_q(X)
\end{equation}
with equality if and only if $p_i=\left[ 1-(1-q)l_i\right]^{\frac{1}{1-q}}$\\
\textbf{Proof.}\\
From Eq(\ref{eqn:Dq}) the generalized KL entropy between two distributions $p$ and $r$ is 
written as 
\begin{eqnarray}
D_q(p\parallel r) & = & \frac{\sum_ip_i^q\left( \ln_qp_i
-\ln_qr_i\right)}{\sum_ip_i^q}=\frac{1-\sum_ip_i^qr_i^{1-q}}{(1-q)\sum_ip_i^q}
\nonumber\\
& = & \frac{1-\sum_ip_i^q}{(1-q)\sum_ip_i^q}-\frac{\sum_ip_i^q(r_i^{1-q}-1)}
{(1-q)\sum_ip_i^q}\label{eqn:Dqpr}.
\end{eqnarray}
If we take the information content associated with probability $r$ as the $i$-th codeword 
length $l_i$, i.e, $-(r_i^{1-q}-1)/(1-q)=l_i$, then the average codeword length can be 
written from Eq(\ref{eqn:Dqpr}) as 
\begin{equation}
\langle L\rangle_q =H_q(X)+D_q(p\parallel r)\label{eqn:LHD}
\end{equation}
Since $D_q(p\parallel r)\ge 0$ for $q>0$ from Theorem 4, we have the theorem. The equality 
holds if and only if $p_i=r_i$. 
$\Box$\\
We note that the relation $-(r_i^{1-q}-1)/(1-q)=l_i$ means that the codeword length 
$l_i$ equals to the information content different from $p$, $l_i=I_q(r)$.
When the equality is realized in the above theorem, we can derive an interesting interpretation on the nonadditivity parameter $q$. The condition of the equality states that the 
probability is expressed as the Tsallis's canonical ensemble like factor in an i-wise manner. 
Then each $l_i$ has the limit in length corresponding to $q$ such as $l_i^{max}=1/(1-q)$.

Since $\log_D \left( \sum_i p_i^q\right) -q\log_D p_i$ obtained by the optimization 
problem is not always equal to an integer, we impose the integer condition on the codewords $\{l_i\}$ by rounding it up as $l_i=\lceil \log_D \left( \sum_i p_i^q\right) -q\log_D p_i \rceil$, 
where $\lceil x\rceil$ denotes the smallest integer $\ge x$\cite{Cover}. Moreover the relation 
$\lceil \log_D \left( \sum_i p_i^q\right) -q\log_D p_i \rceil \ge \log_D 
\left( \sum_i p_i^q\right)-q\log_D p_i$ leads to 
\begin{eqnarray}
\sum_iD^{-\lceil \log_D \left( \sum_i p_i^q\right) -q\log_D p_i \rceil} & \le & 
\sum_iD^{-(\log_D \left( \sum_i p_i^q\right) -q\log_D p_i)}\nonumber\\
& = & \sum_i\frac{p_i^q}{\sum_ip_i^q}=1.
\end{eqnarray}
Hence $\{l_i\}$ satisfies the Kraft inequality. Moreover, we have the following theorem.\\
\textbf{Theorem 6.}\\
The average codeword length assigned by $l_i=\lceil \log_D \left( \sum_i p_i^q\right) 
-q\log_D p_i \rceil$ satisfies 
\begin{equation}
H_q(p)+D_q(p\parallel r)\le \langle L \rangle_q <  H_q(p)+D_q(p\parallel r)+1.
\end{equation}
\textbf{Proof.}\\
The integer codeword lengths satisfies 
\begin{equation}
\log_D \left( \sum_i p_i^q\right) -q\log_D p_i \le l_i < \log_D \left( \sum_i p_i^q\right) 
-q\log_D p_i +1.
\end{equation}
Multiplying by $p_i^q/\sum_ip_i^q$ and summing over $i$ with Eq(\ref{eqn:LHD}) yields 
the theorem. $\Box$\\
This means that the distribution different from the optimal one provokes a correction 
of $D_q(p\parallel r)$ in the average codeword length as does in the case of additive one.\\
We have discussed the properties of the nonadditive entropy in the case of one letter so far. 
Next, let us consider the situation which we transmit a sequence of $n$ letters from 
the source in such a way that each letter is to be generated independently as identically 
distributed random variables according to $p(x)$. Then the average codeword length per letter 
$\langle L_n\rangle_q = \langle l(X_1,\ldots ,X_n)\rangle_q/n$ is bounded as we derived in the 
preceding theorem, 
\begin{eqnarray}
H_q(X_1,\ldots ,X_n) \le \langle l(X_1,\ldots ,X_n)\rangle_q < H_q(X_1,\ldots ,X_n) +1.
\end{eqnarray}
Since we are now considering independently, identically distributed random variables 
$X_1,\ldots ,X_n$, we obtain
\begin{equation}
\frac{\sum_{i=1}^n\left[ 1+(q-1)H_q(X_{i-1},\ldots ,X_1)\right]H_q(X_i)}{n}
\le \langle L_n\rangle_q < \frac{\sum_{i=1}^n\left[ 1+(q-1)H_q(X_{i-1},\ldots ,X_1)\right]
H_q(X_i)}{n}+\frac{1}{n},
\end{equation} 
where we have used Theorem 3. This relation can be considered to be the generalized source 
coding theorem for the finite number of letters. We note that we obtain $H(X) \le 
\langle L_n\rangle < H(X)+1/n$ in the additive limit since $H(X_1,\ldots ,X_n)
=\sum_iH(X_i)=nH(X)$ holds.  
\section{The generalized Fano's inequality}
Fano's inequality is an essential ingredient to prove the converse to the 
channel coding theorem which states that the probability of error that arises 
over a channel is bounded away from zero when the transmission rate exceeds the 
channel capacity.\cite{future}. In the estimation of an original message generated 
from the information source, the original variable $X$ may be estimated as $X'$ 
on the side of a destination. Therefore, 
we introduce the probability of error $P_e=Pr\{ X'\ne X\}$ due to the noise of the channel 
through which the signal is transmitted. With an error random variable $E$ defined as 
\begin{equation}
E = 
\left\{ \begin{array}{rl}
        1 & \mbox{if $X'\ne X$}\\
        0 & \mbox{if $X'=X$}
        \end{array}
\right.\label{eqn:defE}      
\end{equation}
we have the following theorem which is considered to be the generalized(nonadditive 
version) Fano's inequality.\\
\textbf{Theorem 7.} The generalized Fano's inequality\\
\begin{equation}
H_q(X\mid Y)\le H_q(P_e)+\frac{1+(q-1)H_q(E,Y)}{1+(q-1)H_q(Y)}\frac{P_e^q}
{P_e^q+(1-P_e)^q}\frac{1-(|\H|-1)^{1-q}}{(q-1)(|\H|-1)^{1-q}}
\end{equation}
where $|\H|$ denotes the size of the alphabet of the information source.\\
\textbf{Proof.}\\
The proof can be done along the line of the usual Shannon's additive 
case(e.g.\cite{Cover}). Using the corollary Eq(\ref{eqn:Cor}), we have two different 
expressions for $H_q(E,X\mid Y)$,
\begin{equation}
H_q(E,X\mid Y) = H_q(X\mid Y)+\frac{1+(q-1)H_q(X,Y)}{1+(q-1)H_q(Y)}H_q(E\mid X,Y)
\end{equation}
and
\begin{equation}
H_q(E,X\mid Y) = H_q(E\mid Y)+\frac{1+(q-1)H_q(E,Y)}{1+(q-1)H_q(Y)}H_q(X\mid E,Y)
\label{eqn:2ndexp}
\end{equation}
where we have used the corollary by regarding $H_q(E,X\mid Y)$ as $H_q(X,E\mid Y)$ 
in the second expression Eq(\ref{eqn:2ndexp}). We are now considering the following 
situation. That is, we wish to know the genuine 
random variable $Y$ which can be related to the $X$ by the nonadditive conditional 
information content $I_q(y\mid x)$. Hence we calculate $X'$, an estimate of $X$, as 
a function of $Y$ such as $g(Y)$\cite{Cover}. Then we see $H_q(E\mid X,Y)$ becomes $0$ 
since $E$ is a function of $X$ and $Y$ by the definition Eq(\ref{eqn:defE}). Therefore 
the first expression of $H_q(E,X\mid Y)$ reduces to $H_q(X\mid Y)$. On the other hand, 
from the non-negativity property of $\I(E;Y)$ we assumed and from the relation 
$H_q(E)=H_q(P_e)$, we can evaluate $H_q(E\mid Y)$ as $H_q(E\mid Y)\le H_q(E)=
H_q(P_e)$. 
Moreover, $H_q(X\mid E,Y)$ can be written as 
\begin{eqnarray}
H_q(X\mid E,Y) & = & \frac{\sum_E(Pr\{ E\})^qH_q(X\mid Y,E)}
{\sum_E(Pr\{ E\})^q}\nonumber\\
& = & \frac{(1-P_e)^qH_q(X\mid Y,0)+P_e^qH_q(X\mid Y,1)}{P_e^q+(1-P_e)^q}.\label{eqn:HqXEY}
\end{eqnarray}
For $E=0$, $g(Y)$ gives $X$ resulting in $H_q(X\mid Y,0)=0$ and for $E=1$, we have 
upper bound on $H_q(X\mid Y,1)$ by the maximum entropy comprising of the remaining 
outcomes $|\H|-1$,
\begin{equation}
H_q(X\mid Y,1)\le \frac{1-(|\H|-1)^{1-q}}{(1-q)(|\H|-1)^{1-q}}.
\end{equation} 
Then it follows from Eq(\ref{eqn:HqXEY}) that 
\begin{equation}
\left[P_e^q+(1-P_e)^q \right]H_q(X\mid E,Y)\le P_e^q\frac{1-(|\H|-1)^{1-q}}
{(1-q)(|\H|-1)^{1-q}}.
\end{equation}
Combining the above results with Eq(\ref{eqn:2ndexp}), we have the theorem.$\Box$\\
Remark: In the additive limit, we obtain the usual Fano's inequality 
$H(X\mid Y) \le H(P_e)+P_e\ln (|\H|-1)$ in \textit{nat} unit.

\section{Concluding Remarks}

We have attempted to extend the Shannon's additive IT to the nonadditive case by using 
the nonadditive information content Eq(\ref{eqn:NIC}). In developing the nonadditive IT, 
this postulate of the nonadditive information content seems to plausible 
selection in terms of the unification of the meaning of \textit{average} 
throughout the entire theory. As a consequence, the information entropy becomes the 
modified type of Tsallis nonextensive entropy. We have shown that the properties of the 
nonadditive information entropy, conditional entropy and the joint entropy in the form of 
theorem, which are necessary elements to develop IT. These results recover the usual 
Shannon's ones in the additive limit($q\rightarrow 1$). Moreover, the source coding 
theorem can be generalized to the nonadditive case. As we have seen in the theorem 5, 
the nonadditivity of the information content can be regarded that it determines the 
longest codeword we can transmit to the channel. 
Philosophy of the present attempt can be 
positioned as a reverse of Jaynes's pioneering work\cite{Jaynes1,Jaynes2}. 
Jaynes has brought a concept of IT to statistical mechanics in the form of 
maximizing entropy of a system (Jaynes maximum entropy principle). The information 
theoretical approach to statistical mechanics is now considered to be very robust 
in discussing some areas of physics. In turn, we have approached IT 
in a nonadditive way. We believe that the present consideration based on the nonadditive 
information content may trigger some practical future applications in such an area of 
the information processing.

\acknowledgements
The author would like to thank Dr.S.Abe for useful comments at the Yukawa 
Institute for Theoretical Physics, Kyoto, Japan and for suggesting him the nonadditive 
conditional entropy form Eq(\ref{eqn:defcond}).


\begin{references}
\bibitem{check} From the definition, we can easily confirm $H_q(Y\mid X)=H_q(Y)$ if and only if                 $X$ and $Y$ are independent variables. That is, $H_q(Y\mid x)=H_q(Y)$. Moreover,
                since $\langle \left[ 1+(q-1)H_q(Y)\right]^{-1}\rangle^{(X)}_q=
                \langle \displaystyle{\sum_{y\in \Y}}p^q(y)\rangle^{(X)}_q=
                \displaystyle{\sum_{y\in \Y}}p^q(y)$, we obtain $H_q(Y\mid X)=
                \left( 1/\displaystyle{\sum_{y\in \Y}}p^q(y)-1\right)/(q-1)=H_q(Y)$.
\bibitem{qlog} We also note that our definition of $\ln_qx$ corresponds to 
               $-\ln_qx^{-1}$ in the Ref.\cite{RajAbe}.
\bibitem{future} We will report the extension of the theorem elsewhere.
\end{references}
\end{document}